\documentclass[twocolumn,trackchanges]{aastex701}

\usepackage{amsmath,latexsym}    
\usepackage{graphicx}   
\usepackage{verbatim}   
\usepackage{color}      
\usepackage{subfigure}  
\usepackage{hyperref}   
\usepackage{tensor}     
\usepackage{amssymb}
\usepackage{enumerate}	
\usepackage{braket}		
\usepackage{multirow}

\newcommand{\mpcoh}{\,h^{-1}\,{\rm Mpc}}

\newcommand{\bfk}{\boldsymbol{k}}
\newcommand{\bfx}{\boldsymbol{x}}

\newcommand{\bfr}{\boldsymbol{r}}

\newcommand{\PEE}{P_{EE}}

\newcommand{\Pdd}{P_{\delta\delta}}
\newcommand{\PEEl}{P_{EE,\ell}}
\newcommand{\PtEEl}{\wt{P}_{EE,L}}

\newcommand{\xip}{\xi_{+}}
\newcommand{\xim}{\xi_{-}}
\newcommand{\xipm}{\xi_{\pm}}

\newcommand{\xipl}{\xi_{+,\ell}}
\newcommand{\ximl}{\xi_{-,\ell}}
\newcommand{\xipml}{\xi_{\pm,\ell}}

\newcommand{\xitml}{\wt{\xi}_{-,L}}

\newcommand{\be}{\begin{equation}}
\newcommand{\ee}{\end{equation}}

\newcommand{\wt}{\widetilde}
\newcommand{\bftheta}{{\boldsymbol \theta}}

\newcommand{\iip}{\mathrm{II}(+)}
\newcommand{\iim}{\mathrm{II}(-)}

\graphicspath{{./}{figures/}}

\received{July 15, 2025}
\revised{August 3, 2025}
\accepted{August 11, 2025}
\published{August 28, 2025}
\submitjournal{ApJ Letters}

\shortauthors{Okumura}

\begin{document}

\title{
Evidence for Intrinsic Galaxy Alignments in Ellipticity Autocorrelations Out to \\ $100 \mpcoh$ from SDSS Galaxies with DESI Imaging
}

\author[orcid=0000-0002-8942-9772,sname='Okumura']{Teppei Okumura}
\affiliation{Institute of Astronomy and Astrophysics, Academia Sinica, No. 1, Section 4, Roosevelt Road, Taipei 106216, Taiwan}
\affiliation{Kavli Institute for the Physics and Mathematics of the Universe (WPI), UTIAS, The University of Tokyo, Chiba 277-8583, Japan}
\email[show]{tokumura@asiaa.sinica.edu.tw}

\begin{abstract}

Measuring the autocorrelation of galaxy shapes, known as the intrinsic-intrinsic (II) correlation, is important for both cosmology and understanding the formation of massive elliptical galaxies. However, such measurements are significantly more challenging than those of the cross-correlation with galaxy density (GI correlation) due to the much lower signal-to-noise ratio. In this Letter, we report the first observational evidence for large-scale intrinsic alignments measured from the ellipticity autocorrelations, extending out to $100\mpcoh$. From the Sloan Digital Sky Survey (SDSS) and SDSS-III Baryon Oscillation Spectroscopic Survey, we analyze, over the redshift range $0.16\leq z\leq 0.70$, luminous red galaxy, LOWZ, and CMASS galaxy samples, the latter two of which are crossmatched with high-quality Dark Energy Spectrograph Instrument imaging data. By expanding one of the two II correlation functions, $\iim$, in terms of the associated Legendre polynomials, we effectively isolate the line-of-sight projection effects and enhance the signal. The resulting correlation for all three samples exhibits a clear power-law form. We also show that jointly analyzing the two II correlations, $\iip$ and $\iim$, increases the detection significance by $\sim 10\%$, even though both are derived from the same $E$-mode power spectrum. Importantly, this measurement opens a new observational window for probing signals uniquely encoded in shape autocorrelations, such as tensor perturbations from the gravitational waves. Our analysis establishes a practical framework for extracting such effects.
\end{abstract}




\section{Introduction} \label{sec:introduction}

Intrinsic alignments (IAs) of galaxy shapes are a key phenomenon linking large-scale structure with galaxy formation and offering a complementary probe of cosmology.
Originally proposed as a contaminant to weak gravitational lensing measurements, the autocorrelation of galaxy shapes, known as the intrinsic–intrinsic shape (II) correlation, was first studied \citep{Croft:2000,Heavens:2000,Lee:2000,Catelan:2001,Crittenden:2002,Jing:2002}.
\cite{Hirata:2004} later emphasized that the cross correlation between galaxy positions and intrinsic shape (GI correlation) can be an even more serious systematic. 
Because GI correlations are easier to measure observationally, especially using bright galaxies or clusters,  they have become the dominant focus in large-scale IA studies, both theoretically and observationally \citep{Mandelbaum:2006,Hirata:2007,Okumura:2009a,Joachimi:2011,Smargon:2012,Li:2013,Singh:2015,van_Uitert:2017,Desai:2022,Tonegawa:2022,Xu:2023b,Lamman:2024a,Shi:2024,Hervas:2024}.
In contrast, II correlations, which directly probe alignments between galaxy shapes, are significantly noisier and more difficult to detect at large scales. Consequently, only a few studies have succeeded in measuring them with high significance. \cite{Okumura:2009} reported the clearest detection to date, finding an ellipticity autocorrelation signal out to $\sim 30\mpcoh$ using luminous red galaxies (LRGs) from the Sloan Digital Sky Survey (SDSS).

Recently, the potential of IA measurements as cosmological probes has attracted growing attention. In particular, the use of ellipticity correlations in three dimensions has been shown to enable constraints on redshift-space distortions (RSDs) and baryon acoustic oscillations (BAOs), offering both dynamical and geometric information \citep{Chisari:2013,Okumura:2019,Taruya:2020,Okumura:2022,Chuang:2022,Lee:2023,Saga:2023,Shim:2025}. However, observational constraints at large scales continue to rely almost exclusively on GI correlations \citep{Okumura:2023,Kurita:2023,Xu:2023}, due to the difficulty of extracting clean II signals from data.

Despite the challenges, measuring shape autocorrelations remains essential because they carry unique information inaccessible to GI cross correlations. Particularly, shape autocorrelations play a critical role in probing new physics, such as the stochastic gravitational wave (GW) background, through their imprint on galaxy orientations
\citep{Schmidt:2014,Biagetti:2020,Akitsu:2023,Okumura:2024a,Philcox:2024,Saga:2024}.  
Recent work showed that IA signals projected along the line of sight are more naturally expanded using the associated Legendre basis rather than the standard Legendre one, due to the geometric anisotropy of projected shapes \citep{Kurita:2022,Okumura:2024,Singh:2024,Inoue:2025}. This formalism improves the separation of physical signals from noise and enhances interpretability at large scales.

In this Letter, we measure the II correlation functions using various galaxy samples from the SDSS and SDSS-III Baryon Oscillation Spectroscopic Survey (BOSS), covering the redshift range $0.16\leq z\leq 0.70$. To maximize the signal-to-noise ratio, we expand the II correlations in the associated Legendre basis. Following \cite{Xu:2023}, we further improve the shape measurements by crossmatching the BOSS spectroscopic galaxy samples with Dark Energy Spectrograph Instrument (DESI) imaging data. This allows us to report the first observational evidence for intrinsic galaxy alignments in ellipticity autocorrelations out to $100\mpcoh$, a scale comparable to that of the BAO. Our results demonstrate that galaxy shape autocorrelations can be robustly measured at cosmological scales, opening a new window for using IA as a cosmological probe. Throughout this Letter, we adopt a flat $\Lambda$CDM model determined by \cite{Planck-Collaboration:2020}.

\section{SDSS and DESI Galaxy Samples}\label{sec:data}

We analyze the galaxy distribution over $0.16 \leq z \leq 0.70$ using spectroscopic samples from SDSS-II \citep{Eisenstein:2001} and SDSS-III BOSS \citep{Reid:2016}. 
Specifically, we use the LRG sample ($0.16 \leq z \leq 0.47$) from SDSS DR7 and LOWZ ($0.16 \leq z \leq 0.43$) and CMASS ($0.43 \leq z \leq 0.70$) samples from the BOSS DR12, identical to those used in \cite{Okumura:2023}. 

Galaxy ellipticities are defined by two components,
\begin{align}
\left(
\begin{array}{c} \gamma_+ \\ \gamma_\times \end{array}
\right)(\bfx)
 = \frac{1-q^2}{1+q^2} 
\left(
\begin{array}{c} \cos{2\beta} \\ \sin{2\beta}\ \end{array}
\right)
\label{eq:ellip}
\end{align}
where $q$ is the minor-to-major-axis ratio ($0\leq q \leq 1$) and $\beta$ is the position angle of the ellipticity.
To improve the shape measurement quality for the LOWZ and CMASS galaxies, we follow the method of \cite{Xu:2023} and crossmatch these samples with the DESI Legacy Imaging Survey DR9 data \citep{DESI_Collaboration:2016,Dey:2019}.
The DESI imaging covers the entire BOSS footprint and provides imaging data 2 to 3 magnitudes deeper than the SDSS photometry used in target selection, sufficient to robustly determine galaxy orientations.
For the LRG sample, we use the original shape measurements defined by the $25 ~{\rm mag}~{\rm arcsec}^{-2}$ isophote in the $r$ band, which has been shown to be effective for IA analysis. Accordingly, we adopt the same LRG dataset as in \cite{Okumura:2023}.

The total numbers of the LRG, LOWZ, and CMASS used in this study are 105,334, 347,262, and 689,721, respectively.
As in our previous IA analyses, we set the axis ratio to $q=0$ in Equation~(\ref{eq:ellip}) \citep{Okumura:2009a,Okumura:2009,Okumura:2019,Okumura:2020a}. Since our primary interest is in the statistical accuracy of the shape correlations, rather than the amplitude of IA, this simplification does not affect our conclusions.

\begin{figure*}[tb]
\begin{center}
\includegraphics[width = 0.99\textwidth]{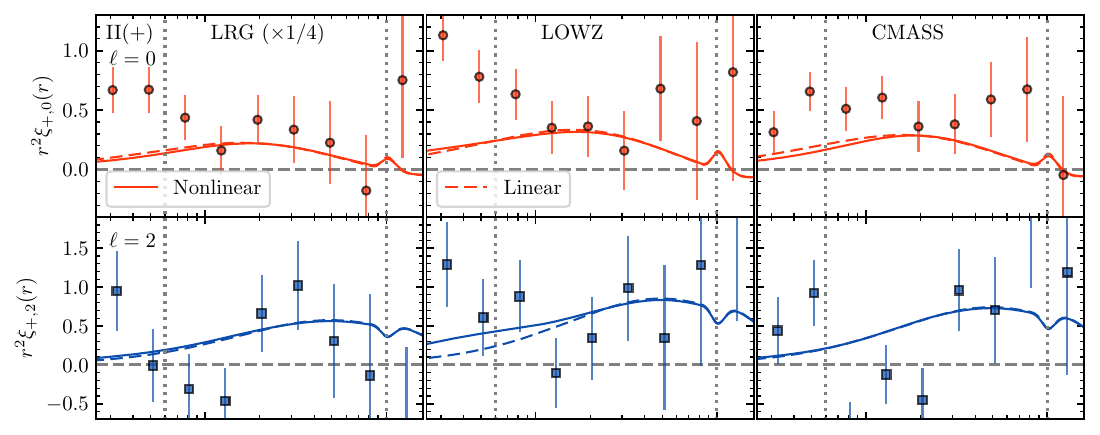}
\includegraphics[width = 0.99\textwidth]{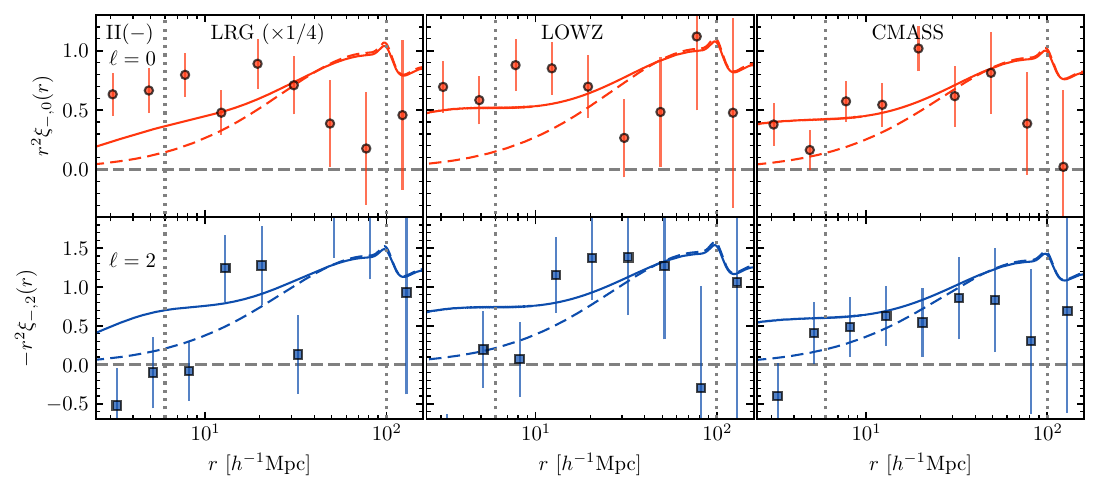}
\caption{
Multipole moments of the intrinsic ellipticity autocorrelation functions expanded in the standard Legendre basis, measured from SDSS LRG (left), LOWZ (middle), and CMASS (right) galaxy samples.
The upper set shows the $\iip$ correlation multipoles, $\xipl$, while the lower set shows the $\iim$ multipoles, $\ximl$, for $\ell=0$ (monopole; circles) and $\ell=2$ (quadrupole; squares). The correlation amplitude for the LRG sample is multiplied by $1/4$. The LRG results are reproduced from \cite{Okumura:2023}, while the LOWZ and CMASS measurements incorporate high-quality DESI imaging data, significantly improving the signal. The solid curves are the best-fit nonlinear model fitted using the data points enclosed by the vertical lines. The dotted curves are the linear predictions. 
} 
\label{fig:xipm}
\end{center}
\end{figure*}


\section{Measurement of Correlation Functions}

We measure two types of intrinsic ellipticity autocorrelations, $\iip$  and $\iim$ correlations 
denoted by $\xip$ and $\xim$, respectively.
These are defined as density-weighted two-point statistics of galaxy ellipticities:
\be
\xipm(\bfr) = \left\langle \left[1+\delta_g(\bfx_1)\right] \left[1+\delta_g(\bfx_2)\right] W_\pm(\bfx_1,\bfx_2)\right\rangle,
\ee
where $W_{\pm}(\bfx_1,\bfx_2) = \gamma_+(\bfx_1)\gamma_+(\bfx_2) \pm
\gamma_\times(\bfx_1)\gamma_\times(\bfx_2)$.  
We estimate $\xipm$ using the pair-based estimator \citep{Mandelbaum:2006},
\begin{align}
&\xi_{\pm}({\bfr}) 
=\frac{S_+S_+ \pm S_\times S_\times}{\mathrm{RR}}\ , 
\end{align}
where $\mathrm{RR}$ is the normalized count of random-random pairs, and $S_+S_+ = \sum_{i\neq j| {\bfr}} {\gamma_+(j|i)\gamma_+(i|j)}$, with $\gamma_+(j|i)$ being the ellipticity of galaxy $j$ measured relative to the direction to galaxy $i$,  and $S_\times S_\times$ is defined similarly. 

Since the IA correlations are anisotropic functions, we bin galaxy pairs in both radial separation $r = |\bfr |$ and the cosine of the angle to the line of sight, $\mu_{\bfr}$. 
We then compute the multipole moments of the correlation functions using the Legendre expansion,
\be
\xipml(r) = \frac{2\ell + 1}{2} \sum_{i} \Delta\mu_{\bfr} ~\xipm(r,\mu_{\bfr,i}){\cal L}_\ell (\mu_{\bfr,i}) ~, \label{eq:multipole}
\ee
where ${\cal L}_\ell$ is the $\ell$th-order Legendre polynomial and $\Delta\mu_{\bfr}=0.1$ is the angular bin size.
Since the geometric factor arising from the projection of the shape field along the line of sight (see Eq.~(\ref{eq:nla_rsd_ii}) below) contributes only to the $\iim$ correlation, it is more naturally expanded in terms of the associated Legendre polynomials, 
\begin{align}
\xitml (r) &= \sum_{i}\Delta\mu_{\bfr} \,  \xim (r,\mu_{\bfr,i})\, \Theta_{L}^{m=4}(\mu_{\bfr,i}) ~,
\label{eq:xim_multipole_associated_legendre}
\end{align}
where $\Theta_{L}^{m}(\mu)$ is the normalized associated Legendre function related to the unnormalized one by
$\Theta_{L}^{m}(\mu) =\sqrt{\frac{2L+1}{2}\frac{(L-m)!}{(L+m)!}}\mathcal{L}_{L}^{m}(\mu_{\bfk})$ and $L\geq m=4$.
We use a tilde to distinguish these coefficients from those expanded in the standard Legendre basis. 
The summation in Eqs.~(\ref{eq:multipole}) and (\ref{eq:xim_multipole_associated_legendre}) is performed over  $-1\leq \mu_{\bfr}\leq 1$.

\begin{figure*}[tb]
\begin{center}
\includegraphics[width = 0.99\textwidth]{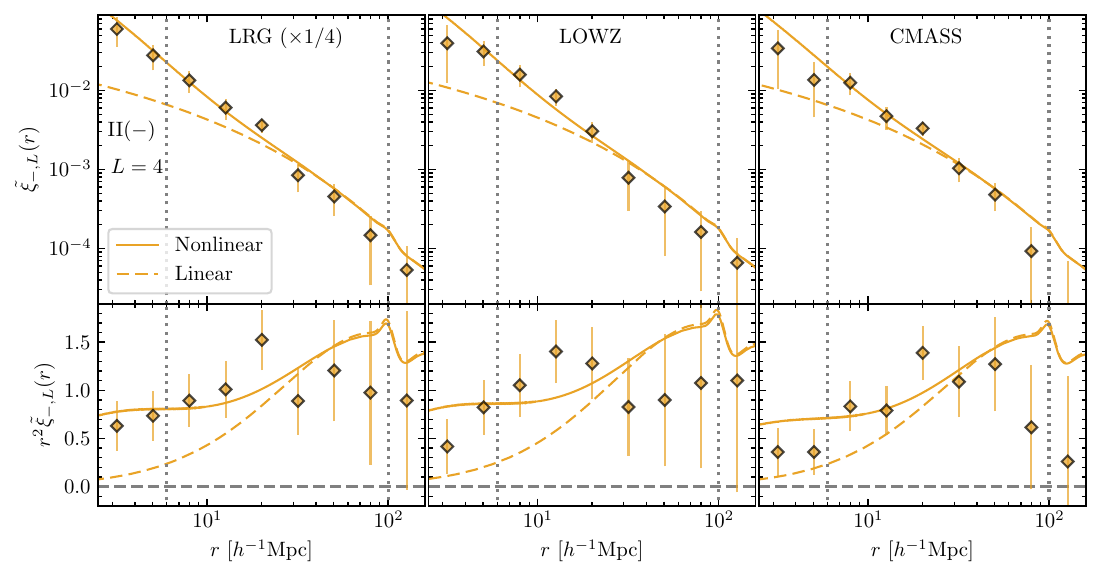}
\caption{Upper panels: similarly to Fig.~\ref{fig:xipm}, but $L=4$ moments of $\iim$ correlation functions of SDSS galaxies expanded by the associated Legendre basis, $\xitml$.  The solid curves are the best-fit nonlinear alignment and RSD models fitted using the data points enclosed by the vertical lines.
Lower panels: same as upper panels, but the vertical axes are multiplied by $r^2$ so that they can be directly compared to the statistics shown in Fig.~\ref{fig:xipm}.
}
\label{fig:xim44}
\end{center}
\end{figure*}

The covariance matrices of all measured statistics, $ \xipl$, $\ximl$, and $\xitml$, are estimated using the jackknife resampling technique. 
While not an unbiased estimator, the jackknife method provides reliable error bars when the dominant noise is shot noise \citep{Mandelbaum:2006}.  
All error bars presented in the following results are the square roots of the diagonal
elements of the covariance matrix.

The first column of Figure~\ref{fig:xipm} shows the multipole moments of the II correlation functions expanded in the standard Legendre polynomials, $\xipl$ (upper set) and $\ximl$ (lower set), measured from the LRG sample.
These results were already presented in \cite{Okumura:2023}. 
The second and third columns show the corresponding measurements for the LOWZ and CMASS galaxies, respectively, using the newly matched DESI imaging data.

In this analysis, we focus only on the two lowest multipoles: the monopole ($\ell=0$) and quadrupole ($\ell=2$), as the hexadecapole ($\ell=4$) is too noisy.
Even for the lower multipoles, it is nontrivial to argue up to which scale the correlation is detected. 
The monopoles of $\iip$ appear nonzero for all three samples up to several tens of $\mpcoh$, while all the quadrupoles remain statistically consistent with zero. 
For the $\iim$ correlation, the monopoles are better measured than those of $\iip$.
The detection for LRG up to $\sim 40\mpcoh$ corresponds to the earlier finding of \citet{Okumura:2009}.
Notably, the LOWZ and CMASS results are significantly improved compared to those in \cite{Okumura:2023}, due to the enhanced imaging depth and quality provided by the DESI data. 
The quadrupole of $\iim$ for CMASS appears nonzero over a relatively large range of scales.  
However, because the signal is distributed over multiple multipoles and the overall measurements remain noisy, these results do not allow a definitive detection based solely on the standard Legendre expansion.

In the upper row of Fig.~\ref{fig:xim44}, we present the $\iim$ correlations expanded in the associated Legendre basis. The lower row shows the same quantities, but with the vertical axis rescaled by $r^2$, allowing direct comparison with the lower set of Figure~\ref{fig:xipm}.
While both figures represent the same underlying correlations, they are shown in different functional bases.
In contrast to the standard Legendre expansion, where the signal is split across several multipoles, the linear-order intrinsic alignment signal is concentrated entirely in the $L = 4$ mode when expanded in the associated Legendre basis.

The resulting measurements exhibit a clear, smooth power-law trend extending to $100\mpcoh$ for all three samples.
This includes the LRG sample, which closely matches that used in \cite{Okumura:2009}, except for a modest extension from SDSS DR6 to DR7. The much clearer detection achieved here is therefore entirely due to the improved basis choice.
Furthermore, the combination of this optimal basis and the deep DESI imaging enables similarly robust detections of $\iim$ autocorrelations for the LOWZ and CMASS samples.
These constitute the first unambiguous measurements of intrinsic shape autocorrelations extending to $100\mpcoh$, across multiple redshift bins.


\section{Analysis}
\label{sec:analysis}

\subsection{Nonlinear Alignment and RSD Model}
\label{sec:la_model}

\begin{table*}[bt!]
\begin{center}
\caption{Constraints on the Amplitude of IA and Reduced $\chi^2$ Values. }
\begin{tabular}{l c c c c c c c c c c}
\hline 
\hline 
& &   &  \multicolumn{2}{c}{LRG} & &  \multicolumn{2}{c}{LOWZ} && \multicolumn{2}{c}{CMASS}\\
\cline{4-5} \cline{7-8} \cline{10-11}
 &  Statistics    & \ dof \ \  &     $-b_K\sigma_8$       &  $\chi_{\min}^2/{\rm dof}$   &  &    $-b_K\sigma_8$       &  $\chi_{\min}^2/{\rm dof}$  & &    $-b_K\sigma_8$       &  $\chi_{\min}^2/{\rm dof}$  \\
\hline
$\iip$ & $\xipl$ & 10 &  $0.1612^{+ 0.0306}_{ - 0.0395}$  & 1.169 & & $0.1037^{+ 0.0141}_{- 0.0167}$ & $0.823$ & & $0.0935^{+ 0.0133}_{ - 0.0157}$ & 2.909 \\
$\iim$ & $\ximl$ & 10 & $0.2401^{+ 0.0173}_{ - 0.0177}$  & $2.553$ & & $0.1208^{+ 0.0101}_{- 0.0107}$ & $1.584$ & & $0.1176^{+ 0.0082}_{ - 0.0085}$ & 1.092 \\
$\iim$ & $\xitml$ &4& $0.2358^{+ 0.0170}_{- 0.0174}$  & $1.079$ & & $0.1198^{+ 0.0102}_{- 0.0109}$ & $1.391$ & & $0.1174^{ + 0.0083}_{ - 0.0085}$ & 1.109 \\
$\iip +\iim$ \ \ &  $\xipl + \ximl$ &22& $0.2131^{ + 0.0135}_{ - 0.0145}$  & $2.096$ & & $0.1159^{+ 0.0080}_{ - 0.0088}$ & $1.138$ & & $0.1087^{+ 0.0069}_{ - 0.0073}$ & 2.224 \\
$\iip+\iim$ \ \ &$\xipl + \xitml$ &16& $0.2107^{ + 0.0138}_{ - 0.0148}$  & $1.293$ & & $0.1152^{+ 0.0081}_{- 0.0087}$ & $0.893$ & & $0.1083^{ + 0.0070}_{ - 0.0075}$ & 2.604 \\
\hline
\end{tabular}
\label{tab:statistics}
\end{center}
\end{table*}

To model the IA statistics, we consider the nonlinear alignment (NLA) model, which assumes a linear relation between the intrinsic ellipticity and underlying nonlinear tidal field, $\left(\frac{\nabla_i\nabla_j}{\nabla^{2}}\right)\delta_m(\bfx) $
\citep{Catelan:2001,Hirata:2004,Bridle:2007}, In Fourier
space, the ellipticity projected along the line of sight ($z$-axis) is
given by
\begin{align}
\left(
\begin{array}{c} \gamma_+ \\ \gamma_\times \end{array}
\right) (\bfk)=b_K
\left(
\begin{array}{c} 
    (k_{x}^{2}-k_{y}^{2})/k^2 \\
    2k_{x}k_{y}/k^2 
\end{array}
\right)
\delta_m(\bfk),
\end{align}
where 
$\delta_m$ is the nonlinear matter density field and
$b_{K}$ is the redshift-dependent coefficient of the IA, which we refer to as the shape bias.\footnote{The shape bias parameter is related to $\wt{C}_1$ in \cite{Okumura:2020} by $b_K=-\wt{C}_1$.}
Furthermore, the redshift-space shape field is multiplied by the damping function due to the Finger-of-God (FoG) effect.
We then define $E$-/$B$-modes, $\gamma_{(E,B)}$, which are the
rotation-invariant decomposition of the ellipticity field
\citep{Crittenden:2002},
\begin{align}
\gamma_E(\bfk) + i\gamma_B(\bfk) = e^{-2i\phi_k}\left[  \gamma_{+}(\bfk)+i\gamma_{\times}(\bfk) \right],
\label{eq:gamma_E_deltam}
\end{align}
where $\phi_k$ is the azimuthal angle of the wavevector projected on the celestial sphere.
Adopting the NLA model, the autopower spectrum of the $E$-mode in redshift space is given by \citep{Okumura:2024}, 
\begin{align}
P_{EE}(\bfk) &= b_K^2 (1-\mu_{\bfk}^2)^2 P_{\delta\delta}(k)
D_{\rm FoG}^2(k\mu_{\bfk} \sigma_v)~, \label{eq:nla_rsd_ii} 
\end{align}
where $k=|\bfk|$, $\mu_{\bfk}$ is the direction cosine between the
line of sight and $\bfk$, and 
$\Pdd$ is the nonlinear matter power spectrum computed by the revised \texttt{Halofit} \citep{Takahashi:2012}. 
Since the FoG effect enters IA power spectra in the same way as the conventional galaxy power spectrum, we adopt a simple Gaussian function to characterize the FoG damping, $D_{\rm
  FoG}(k\mu_{\bfk}\sigma_v)=\exp{\left(-k^2\mu_{\bfk}^2\sigma_v^2/2\right)}$, with $\sigma_v$ the nonlinear velocity dispersion parameter.
Since $P_{\delta\delta}$ is proportional to the square of the normalization
parameter of the density fluctuation, $\sigma_8^2(z)$, free parameters for this model are $\bftheta=(b_K\sigma_8, \sigma_v)$.

To take into account the line-of-sight anisotropy in Equation~(\ref{eq:nla_rsd_ii}), we use the multipole expansion in terms of the standard or associated Legendre polynomials \citep{Kurita:2022,Okumura:2024},
\begin{align}
\PEE(\bfk)&=\sum_{\ell}\PEEl(k){\cal L}_{\ell}(\mu_{\bfk}) \label{eq:coeff_peel}\\
&=\sum_{L\geq 4}\PtEEl(k)\Theta_{L}^{m=4}(\mu_{\bfk})~. 
\end{align}
The $\iip$ and $\iim$ correlation functions are expanded in terms of the standard and associated Legendre bases, respectively, and given by the inverse Hankel transform,
\begin{align}
&\xipl(r)=i^\ell \int \frac{k^2 dk}{2\pi^2}j_\ell(kr)\PEEl(k)~, \label{eq:xipls} \\
&\xitml (r) =i^L \int \frac{k^2 dk}{2\pi^2}j_L(kr)\PtEEl(k)~, \label{eq:xitmls} \end{align}
where $j_\ell$ is the spherical Bessel function. 
Converting the $\iim$ correlation function from the associated to the standard Legendre bases can be performed via the linear transform,
\begin{align}
\ximl(r) &= \sum_{L}A_{\ell L}\, \xitml(r) ~, \label{eq:ximl}
\end{align}
where the coefficients $A_{\ell L}$ are constants and provided in Appendix \ref{sec:additional_formulas}.
In the linear-theory limit, $\sigma_v \to 0$ and thus $D_{\rm FoG}=1$. Hence,
Equation~(\ref{eq:nla_rsd_ii}) converges to the linear-theory formulas derived
in \cite{Okumura:2020} \citep[see also Appendix C of][]{Okumura:2024}.

\begin{figure}[tb]
\begin{center}
\includegraphics[width = 0.495\textwidth]{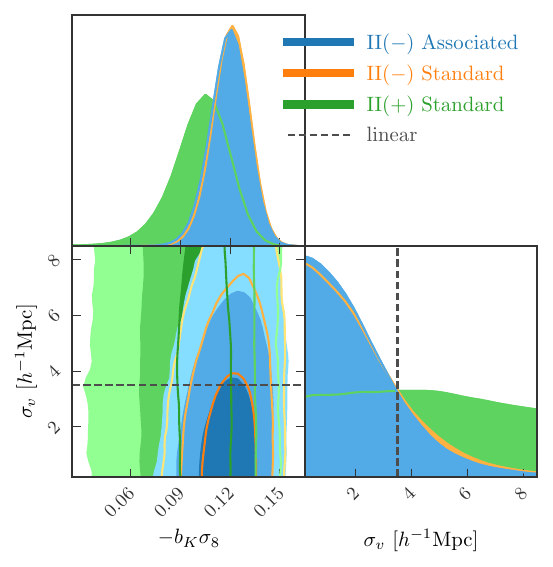}
\caption{Constraints on $(b_K\sigma_8,\sigma_v)$ from II correlation multipoles of LOWZ sample at $6\leq r\leq100\mpcoh$. The contours show the $68\%$, $95\%$, and $99.7\%$ C.~L. from inward. The dashed line represents the linear-theory prediction for velocity dispersion.}
\label{fig:parameters}
\end{center}
\end{figure}


\subsection{Comparison with Models}\label{sec:result}

We perform the likelihood analysis to quantitatively evaluate the detection significance of II autocorrelations for the three SDSS galaxy samples. 
We compute the $\chi^2$ statistics for each measured statistic of $\xipl,\ximl,\xitml$.
We use the $\ell=0,2$ multipoles for $\xipl$ and $\ximl$ and the $L=4$ multipole for $\xitml$, all of which were shown in Figures~\ref{fig:xipm} and \ref{fig:xim44}.

We use the full covariance matrix to account for small correlations between different separation bins and between different statistics, although the II correlations are dominated by shot noise and thus primarily by the diagonal elements \citep[see Fig.~3 of ][]{Chuang:2022}. 
We have two fitting parameters, $\bftheta=(b_K\sigma_8,\sigma_v)$. 
The analysis is performed over the scales adopted, 
$6\leq r \leq 100~[\mpcoh]$. 
Namely, we have 6 data points for each of the multipoles. 
See Table \ref{tab:statistics} for the degree of freedom for each analysis.
The data points used for the analysis are enclosed by the
vertical lines in Figure~\ref{fig:xim44}.

Figure~\ref{fig:parameters} shows the parameter constraints obtained from the LOWZ sample. 
The results for the LRG and CMASS samples are not shown here because the trend of the constraints and their interpretation are very similar to the LOWZ result. 
The green, orange, and blue contours are results with $\xipl,\ximl$, and $\xitml$, respectively. 
The best-fitting values of $b_K\sigma_8$ and the reduced $\chi^2$ are shown in Table \ref{tab:statistics}.
Note that, because we set $q=0$ in Equation (\ref{eq:ellip}), the definition of $b_K$ used here differs from that in the literature, and values are not directly comparable.

As demonstrated in \cite{Okumura:2024}, the $\iip$ correlation is not sensitive to the FoG effect, and the best-fit $b_K$ value is lower than that from the $\iim$ correlation. Besides, $\xipl$ are poorly fit by our model for the CMASS sample. By excluding the small-scale data of the quadrupole and adopting $r_{\min}=25\mpcoh$, the reduced $\chi^2_{\min}$ is converged to $\chi^2_{\min}/\mathrm{dof}\sim1$.
Our model incorporates nonlinear RSD but retains the linear shape bias model, resulting in an overprediction of the amplitude of the $\iip$ correlation and a poor fit to the data on small scales. The nonlinear shape bias beyond the NLA model has been extensively studied 
\citep{Vlah:2020,Akitsu:2021,Akitsu:2023a,Bakx:2023,Chen:2024,Matsubara:2024}. Accounting for such higher-order shape bias effects will enable us to predict the $\iip$ correlation better and to use the measured IA correlation functions down to
smaller scales, which will further enhance the science return from IA of galaxies.

We obtain similar constraints on $b_K$ from $\ximl$ and $\xitml$, which is expected because they are the same statistics but expanded by different bases \citep[see also][]{Inoue:2025}.
This demonstrates that the linear-level signal of the $\iim$ correlation, when expanded by the associated Legendre basis, is well compressed into the single $L=4$ multipole.
Thus, this allows us to emphasize the first clear evidence of the ellipticity autocorrelation up to $\sim 100\mpcoh$ via the appropriate choice of the basis and using the high-quality imaging data.
Note that our model of the $\iim$ correlation, when expanded in the associated Legendre basis, provides a good $\chi^2$ fit to all three samples.

Finally, let us investigate the full signals of the ellipticity correlations, namely, jointly analyze the $\iip$ and $\iim$ correlation functions. 
Since the latter has a flexibility of choice for the basis, we consider the two data vectors $(\xipl,\xi_{+,\ell'})$ and $(\xipl,\wt{\xi}_{-,4})$ with $\ell,\ell'=\{0,2\}$ and construct the corresponding full covariance matrix.
The constraints are shown in Table \ref{tab:statistics}. 
As seen in Equations~(\ref{eq:coeff_peel})--(\ref{eq:xitmls}), all the ellipticity autocorrelation functions are derived from $\PEE$. Nevertheless, combining them leads to $\sim10\%$ improvement in the detectability compared to the $\iim$-only constraints. Even in this case, the choice of the associated Legendre basis for the $\iim$ correlation provides almost the equivalent of the case of the standard Legendre basis that requires the measurement of the multipoles at least $\ell=0$ and 2.


\section{conclusions} \label{sec:conclusion}

We have presented the first observational evidence for IA of galaxies in ellipticity autocorrelation functions extending to $100\mpcoh$, using the LRG, LOWZ, and CMASS galaxy samples from SDSS and SDSS-III BOSS over the redshift range $0.16\leq z\leq 0.70$.
While such autocorrelations are significantly more difficult to measure than the cross correlations with galaxy density fields, they contain unique physical information, including potential imprints of GWs, which are inaccessible to GI correlations.

To mitigate the projection effects of galaxy images along the line of sight, we expand the 
$\iim$ correlation in terms of associated Legendre polynomials, which isolate the linear-order information into a single multipole. 
This basis choice, combined with high-quality DESI imaging data, yields a clear power-law correlation extending to scales comparable to the BAO features. 
The results represent the first unambiguous evidence of large-scale shape autocorrelations across multiple redshift bins.
Although the $\iip$ correlation is noisier, combining it with $\iim$ leads to a $\sim 10\%$ improvement in detection significance. 

Our analysis establishes a robust observational foundation for cosmological applications of ellipticity autocorrelations. These include future efforts to constrain the stochastic GW background as well as the expansion history of the Universe. In particular, since BAO signatures are also encoded in IA statistics \citep{Okumura:2019,Xu:2023}, combining IA measurements with galaxy clustering promises improved geometric constraints. Joint constraints on growth and expansion rates with galaxy IA will be presented in our upcoming Letter \citep{Okumura:2025}.

When we were finalizing the work based on the SDSS spectroscopy data and limited to the redshift of $0.16\leq z \leq 0.70$, the DESI DR1 data were publicly released \citep{DESI_Collaboration:2025}. 
They naturally extend this analysis over a wider redshift range. 
Additionally, optimizing the IA signal through luminosity weighting to brighter galaxies \citep[e.g.,][]{Seljak:2009a} may further enhance sensitivity. 
These directions will be pursued in future work.

\begin{acknowledgments}
T.O. thanks Toshiki Kurita for the discussion on handling the DESI imaging data. 
T.O. acknowledges support from the Taiwan National Science and Technology Council under grant 
Nos. NSTC 112-2112-M-001-034-,
NSTC 113-2112-M-001-011- and
NSTC 114-2112-M-001-004-, and the Academia Sinica Investigator Project grant No. AS-IV-114-M03 for the period of 2025–2029.  
Funding for SDSS-III has been provided by the Alfred P. Sloan Foundation, the Participating Institutions, the National Science Foundation, and the US Department of Energy Office of Science. The SDSS-III website is http://www.sdss3.org/.
\end{acknowledgments}

\appendix


\section{Multipole Expansion of Power Spectra and Correlation Functions}\label{sec:additional_formulas}

In this Appendix, we provide some formulas used to compute the theoretical models of the ellipticity correlation functions but not presented in the body text. Full formulas, including higher-order multipoles, are presented in \cite{Okumura:2024}

Adopting a simple Gaussian function to characterize the FoG damping, all the multipole moments of the power
spectra are expressed in a factorized form with the
angular dependence encoded by
$p^{(n)}(\alpha)\equiv \int^1_{-1} d\mu_{\bfk} \mu_{\bfk}^{2n} e^{-\alpha\mu_{\bfk}^2}
=\gamma(1/2+n,\alpha)\alpha^{-(1/2+n)}$.
The multipole moments of the $EE$ power spectrum expanded in terms of the standard and associated Legendre polynomials are expressed as
\begin{align}
\PEEl(k) 
&= b_K^2 \mathcal{Q}_{EE,\ell} (\alpha)\Pdd(k), 
\qquad\qquad
\PtEEl(k)
= b_K^2 \mathcal{\wt{Q}}_{EE,L} (\alpha)\Pdd(k), \label{eq:PtEEls}
\end{align}
where $L\geq m=4$. 
While the former with $\ell=0,2,4$ contain the linear-order contributions, we show the formulas only up to $\ell=2$ because the hexadecapole was not used in this work.
In the latter, only the lowest-order coefficient with $L=4$, $\mathcal{\wt{Q}}_{EE,4}$, contains the contribution in linear theory.
Using $p^{(n)}$, these coefficients are given by
\begin{align}
\mathcal{Q}_{EE,0}(k) &= \frac12\left[p^{(0)}(\alpha) - 2p^{(1)}(\alpha) + p^{(2)}(\alpha)\right],  \label{eq:QEE0}\\
\mathcal{Q}_{EE,2}(k) &= -\frac54 \left[p^{(0)}(\alpha) - 5p^{(1)}(\alpha) + 7 p^{(2)}(\alpha) 
-3 p^{(3)}(\alpha) \right]. \label{eq:QEE2}
\\
\mathcal{\wt{Q}}_{EE,4}(\alpha)& =\frac{3 \sqrt{35}}{16}\left[ p^{(0)}(\alpha) -4p^{(1)}(\alpha) +6p^{(2)}(\alpha)
-4p^{(3)}(\alpha)+p^{(4)}(\alpha)\right].
\end{align}
Substituting Equations~(\ref{eq:PtEEls}) into Equations~(\ref{eq:xipls}) and (\ref{eq:xitmls}) gives our predictions for $\xipl$ and $\xitml$. 

Finally, to compute $\ximl$, the coefficients in Eq.~(\ref{eq:ximl}) are given as
\begin{align}
\xi_{-,0}(r) = 
& \frac{\sqrt{35}}{10} \wt{\xi}_{-,4}(r)-\frac{\sqrt{91}}{35}\wt{\xi}_{-,6}(r)+\frac{\sqrt{1309}}{210} \wt{\xi}_{-,8}(r)
-\frac{2\sqrt{5005}}{1155}\wt{\xi}_{-,10}(r) +\frac{25\sqrt{2002}}{12012} \wt{\xi}_{-,12}(r)
+\cdots \, ,
\label{xim0s} \\
\xi_{-,2}(r) = &-\frac{\sqrt{35}}{7} \wt{\xi}_{-,4}(r)-\frac{\sqrt{91}}{14}\wt{\xi}_{-,6}(r)+\frac{4\sqrt{1309}}{231} \wt{\xi}_{-,8}(r)
 -\frac{43\sqrt{5005}}{6006}\wt{\xi}_{-,10}(r)
+\frac{5\sqrt{2002}}{546} \wt{\xi}_{-,12}(r)+\cdots \, .
\label{xim2s}
\end{align}

\bibliographystyle{apj}
\bibliography{ms.bbl}

\end{document}